\documentclass[12pt]{article} 
\usepackage{amsmath}
\usepackage{amssymb}
\usepackage{euscript}
  
\usepackage{epsfig}  
  
\usepackage{cite}  
  
\usepackage{alltt}  
  
\usepackage{epic}  
  
\begin{document}     
\allowdisplaybreaks

\begin{titlepage}  
  
\begin{flushright}  
{\bf   
SLAC-PUB-9154 \\  
CERN-TH/2002-054\\March 2002}  
\end{flushright}  
  
\hspace{1 mm}  
\vspace{1 cm}  
\begin{center}{\bf\Large {\boldmath Measuring the Higgs boson's parity  }}\end{center}  
\begin{center}{\bf\Large using    
{\boldmath $\tau \to \rho \nu$} $^{*}$ } \end{center}  
\begin{center}{\bf\Large }\end{center}  
\vspace{0.3 cm}  
\begin{center}  
  {\large\bf G. R. Bower$^{a}$, T. Pierzcha\l a$^{b}$,  Z. W\c{a}s$^{c,d}$} ~{\large \bf and}~ {\large\bf   M. Worek$^{b}$ }  
\vspace{0.3 cm}  
\\  
{\em $^a$ Stanford Linear Accelerator Center, Stanford University\\  
 Menlo Park, CA 94025, USA.}\\  
{\em $^b$ Institute of Physics, University of Silesia\\ Uniwersytecka 4,   
40-007 Katowice, Poland.}\\  
{\em $^c$Institute of Nuclear Physics\\  
         Kawiory 26a, 30-055 Cracow, Poland.}\\  
{\em $^d$CERN, Theory Division, CH-1211 Geneva 23, Switzerland.}\\  
  
\vspace{0.5 cm}  
  
\end{center}  
  
\vspace{1mm}  
\begin{abstract}  
We present a very promising method for a  
measurement of the Higgs boson parity using the   
$H/A \to \tau^{+}\tau^{-} \to \rho^{+}\bar{\nu}_{\tau}\rho^{-}\nu_{\tau}  
\to \pi^{+}\pi^0\bar{\nu}_{\tau}\pi^{-}\pi^0\nu_{\tau}$   
decay chain. The method is both model independent and independent of  
the Higgs production mechanism. Angular   
distributions of the $\tau$ decay products which are sensitive to the Higgs boson  
parity are defined and are found to be measurable using typical properties of a  
future detector for an $e^+e^-$ linear collider. The prospects  
for the measurement of the parity of a Higgs boson with a mass of 120 GeV are  
quantified for the case of $e^+e^-$ collisons of 500 GeV  
center of mass energy with an integrated luminosity of $500 fb^{-1}$. The Standard Model   
Higgsstrahlung production process is used as an example.  
\end{abstract}  
\vspace{2mm}  
\centerline{\it Phys. Lett. B in print}
\vspace{2mm}  
\footnoterule  
\noindent  
{\footnotesize  
\begin{itemize}  
\item[${*}$]  
This work is partly supported by  
the Polish State Committee for Scientific Research   
(KBN) grants Nos 5P03B10121, 2P03B00122  
and the European Community's Human Potential  
Programme under contract HPRN-CT-2000-00149 Physics at Colliders.  
This work is also supported in part by the United States  
Department of Energy contract DE-AC03-76SF00515.  
\end{itemize}  
}  
  
\end{titlepage}  
\section{Introduction}  
If the Higgs mechanism   
is realized in nature,  
a complete determination of its properties is among the
central tasks for future colliders. In particular, determining    
the ${\cal CP}$ properties of the Higgs boson is one of   
the important goals of a future $e^{+}e^{-}$ linear collider operating at a  
center-of-mass energy between $350$$-$$1000$   
GeV \cite{Abe:2001gc,:2001ve,TESLA}. Different models, such as   
the Standard Model (${\cal SM}$)   
, the general Two Higgs Doublet   
Models (2${\cal HDM}$),   
the Minimal Supersymmetric   
Standard Model (${\cal MSSM}$),   
and others  
predict different properties  
for the details of the Higgs boson signature. It is of great   
importance to verify that, whatever scenario is realized in nature,  
the scientific program of a linear collider will be able to distinguish   
among these models.  
  
Several methods have been proposed to   
distinguish between a scalar (${\cal J^{PC}}=0^{++}$) and a pseudoscalar   
(${\cal J^{PC}}=0^{-+}$) spin zero Higgs  
 particle \cite{Kramer:1994jn,Barger1994,Hagiwara1994,Hagiwara2000,Grzadkowski:1995rx,Boe,Skjold}.  
In this study we investigate the case of a Higgs boson which is light   
enough that the $W^+W^-$ decay channel remains closed.   
Then the most promising decay channel for the model independent parity determination    
is $H/A \to \tau^+\tau^-$.   
It was previously proposed \cite{Kramer:1994jn} that the best   
 $\tau$ decay channel for parity determination would be   
$\tau^\pm\to \pi^\pm \nu$. The weakness of this method is that  
the Higgs boson rest frame needs to be precisely reconstructed.  
  
To address resolution issues it is necessary to perform Monte Carlo studies  
where the significant details of theoretical effects and detector conditions can   
be included.  
In Ref. \cite{Was2002} we have extended the algorithm   
of Refs. \cite{Pierzchala:2001gc,Golonka:2002iu}   
for generating Higgs decay (independently from its production mechanism)   
to the pair of   
$\tau$-leptons including the complete spin correlation matrix used in   
the subsequent decay   
of the $\tau$ leptons. The reaction   
chain $e^+e^- \to Z (H/A)$,    
$H/A \to \tau^{+}\tau^{-}$;  
$\tau^{\pm}\to \pi^{\pm}\bar{\nu}_{\tau}({\nu}_{\tau})$ was   
studied. It was found that even small effects of smearing  
seriously deteriorate the measurement resolution.   
An independent study \cite{Bower}, using  
the {\tt PANDORA} Monte Carlo generator of Ref. \cite{Pandora} confirms that result.   
This leaves no doubt that if we want to use   
the Higgs boson $\tau$ decay channel for the measurement of the Higgs parity, a better technique  
is required.  
  
In this paper we continue to investigate Higgs decay  
into a $\tau^{+}\tau^{-}$ pair.  
We extend our study to the    
$H \to \tau^+\tau^-$ decay with   
$\tau^{\pm} \to \rho^{\pm}\bar{\nu}_{\tau}(\nu_{\tau})$ ({\it i.e.} the channel with  
the largest branching ratio) and then    
$\rho^{\pm}\to\pi^{\pm}\pi^{0}$.   
Spin correlations are distributed over {\it three} levels  
of the decay chain. Complicated geometrical distributions need to be defined.  
We have found the Monte Carlo method particularly useful already at the level  
of defining the observables.   
All of the main results are cross-checked using two independent analyses  
based on an extension of universal interface (Ref.~\cite{Was2002}) for {\tt TAUOLA},   
and with the {\tt PANDORA}   
Monte Carlo generator~\cite{Pandora} which is interfaced to   
{\tt PYTHIA 6.1} \cite{Pythia}  
for fragmentation and decay processes.

The rest of the paper is organized as follows. The theoretical considerations   
which are used to understand the decay chain   
of the Higgs particle are  explained in section 2 where we also   
give some details of the {\tt TAUOLA} Monte Carlo simulation.  
In section 3 we define the observable we  
use to distinguish between the  
scalar and pseudoscalar Higgs boson.  
In section 4 we list our assumptions on smearing  
and we discuss imposed cuts, detector effects   
and necessary adaptations introduced into our observables.  
Main numerical results are also given in this section.  
Section 5, a summary, closes the paper.

\section{Determination of the ${\cal CP}$ quantum numbers of the Higgs boson}  
  
The $H/A$ parity information must be extracted from  
the correlations between $\tau^{+}$ and $\tau^{-}$ spin components which are  
further reflected in correlations between    
the $\tau$ decay products in the plane transverse to the   
$\tau^{+}\tau^{-}$ axes. This is because  the   
decay probability, see \cite{Kramer:1994jn},  
\begin{equation}  
\Gamma(H/A\to \tau^{+}\tau^{-}) \sim 1-s^{\tau^{+}}_{\parallel}  
s^{\tau^{-}}_{\parallel}\pm s^{\tau^{+}}_{\perp}s^{\tau^{-}}_{\perp}  
\label{densi}  
\end{equation}   
is sensitive to the $\tau^\pm$ polarization   
vectors $s^{\tau^{-}}$ and $s^{\tau^{+}}$  (defined    
in their respective rest frames).    
The symbols ${\parallel}$/${\perp}$ denote components parallel/transverse   
to the Higgs boson momentum as seen from the respective $\tau^\pm$  rest frames.  
This suggests that the experimentally clean $\tau^{+}\tau^{-}$ final state  
may be the proper instrument to study the parity of the Higgs boson.  
  
The spin of the $\tau$ lepton is not directly observable but it manifests itself  
in the distributions of its decay products. Depending on the decay channel, the  
polarimetric strength is different. The first Monte Carlo program  
for $e^+e^-$ colliders where  
the density matrix of the $\tau$ lepton pair was used was {\tt KORALB}   
\cite{jadach-was:1984,koralb:1985}.   
Let us recall some basic properties of that solution, which we have adopted to  
the case of the Higgs boson decay in Ref. \cite{Was2002}.  
The algorithm is organized in two steps. In the first step, the $\tau$ lepton pair  
is generated and the $\tau$ leptons are decayed in their respective   
rest frames as if there   
were no spin effects at all. In the second step, the spin weight is calculated  
and rejection is performed. If the event is rejected, only   
the generation of the $\tau$ lepton decays is repeated.  
The spin weight is given by the following formula  
\begin{equation}  
  wt= {1 \over 4} \Bigl( 1 + \sum_{i,j=1,3} R_{i,j} h^i h^j \Bigr)  
\end{equation}  
where, as a consequence of formula (\ref{densi}), the components   
$R_{3,3} =-1$, $R_{1,1} =\pm 1$, $R_{2,2} =\pm 1$  
(respectively for scalar and pseudoscalar) and all other components are  
zero\footnote{ See Ref.~\cite{Was2002} for detailed definition of the   
quantization frames   
used for the spins of the $\tau^+$ and $\tau^-$.}.  
  
In the following, we focus on the    
$\tau^{\pm} \to \rho^{\pm}  \nu$   
decay channel. It is interesting because it has, by   
far, the largest  
branching ratio (25\%). However, in comparison to  $\tau^{\pm} \to  
\pi^{\pm} \nu$ decay, its polarimetric force is more than a factor of 2 smaller.  
It  was found, see {\it e.g.} \cite{Harton:1995dj,Nelson:1995vt},   
that in many cases   
this can be improved if information on  
the $\rho$ decay products, {\it i.e.} on details of  
the decay  $\tau^{\pm} \to \pi^{\pm} \pi^0 \nu$, are used.  
This is of no surprise because the polarimetric vector is given by the formula  
\begin{equation}  
h^i =  {\cal N} \Bigl( 2(q\cdot N)  q^i -q^2  N^i \Bigr)  
\end{equation}  
 where ${\cal N}$  is a normalization function,  $q$  is the difference   
of the $\pi^\pm $ and $\pi^0$ four-momenta and $N$ is the four-momentum of the   
$\tau$ neutrino (all defined in the $\tau$ rest frame)   
see, {\it e.g.} \cite{Jadach:1990mz}.   
Obviously, any   
control on the vector $q$ can be advantageous. It is of interest to note that  
in the $\tau$ lepton rest frame, when $m_{\pi^{\pm}}=m_{\pi^{0}}$ is   
assumed, the term   
\begin{equation}  
 q\cdot N = (E_{\pi^\pm} - E_{\pi^0}) m_\tau.  
\label{formula}  
\end{equation}  
Thus, to exploit this part of the polarimetric vector,   
we need to have some handle   
on the difference of the $\pi^\pm$ and $\pi^0$ energies in their respective  
$\tau$ leptons rest frames. Otherwise, the effect of this part of the   
polarimetric vector cancels  
out and one is left with the part proportional to the $\rho$ (equivalently   
$\nu$) momentum. Without using the energy difference we would arrive at   
a case nearly identical to the one where the $\tau$ decays to a  
single $\pi$ except with the disadvantage that the coefficient   
diminishes the spin effects by more than a factor of 2.  
  
Already from this preliminary discussion we realize that the appropriate   
observable must rely on a distribution constructed out of at least 4 four-vectors.  
The Monte Carlo method is already essential at the level   
of designing the observable.  
Our search for improvements of the results obtained in   
\cite{Kramer:1994jn,Was2002}  
began with a study of  acollinearity distributions defined in   
the Higgs boson rest frame,  
when instead of the $\tau \to \pi \nu$ decay mode the mode $\tau \to \rho\nu$ was used.  
The difference between the scalar and pseudoscalar Higgs particle  
 which was still visible for $\tau \to \pi \nu$ turned out to be    
practically invisible  
in the case of $\tau \to \rho\nu$  
once detector smearings were introduced.  
We found these results rather   
discouraging,  
and we have  turned our attention to another possibility. As inspiration we have   
used two methods, one   
used at LEP 1 for the measurement of the $\tau$ polarization and another one, proposed for low energy $e^+e^-$ colliders, see {\it e.g.} \cite{Harton:1995dj,Nelson:1995vt}.

\subsection{The Monte Carlo}  
For the following discussion all the Monte Carlo samples have been generated with the {\tt TAUOLA}  
library \cite{Jadach:1990mz,Jezabek:1991qp,Jadach:1993hs}.  
The {\tt PHOTOS} \cite{Barberio:1990ms,Barberio:1994qi}  
Monte Carlo program could be used for generating radiative corrections   
in the decays  
of the Higgs boson and $\tau$ leptons, but it was switched off.  
For the production of the $\tau$ lepton pairs the Monte Carlo program   
{\tt PYTHIA 6.1} \cite{Pythia}  
is used.\footnote{ It was shown that the interface can work as well in   
the same manner with the   
{\tt HERWIG} \cite{HERWIG} generator.} The production process    
$e^{+}e^{-}\to ZH \to \mu^{+}\mu^{-} (q \bar q) H$ has been   
chosen with a Higgs boson mass of $120$ $GeV$ and a   
center-of-mass energy of $500$ $GeV$. The effects of initial state   
bremsstrahlung were included in the {\tt PYTHIA} generation.  
  
For  the $\tau$ lepton pair decay   
with full spin effects included in the   
$H \to \tau^+\tau^-$, $\tau^{\pm} \to \rho^{\pm}\bar{\nu}_{\tau}(\nu_{\tau})$,  
$\rho^{\pm}\to\pi^{\pm}\pi^{0}$  
chain, the  
interface explained in  Ref.~\cite{Was2002} was used.  
It is an extended version of the   
standard universal   
interface of Refs.~\cite{Pierzchala:2001gc,Golonka:2002iu}.  For the sake  
of confidence we have confirmed  all numerical results presented in this paper  
with the second simulation   
using the {\tt PANDORA} Monte Carlo generator~\cite{Pandora}.

  
\section{The acoplanarity of the $\rho^+$ and $\rho^-$ decay products}

In this section we advocate a new observable where we   
ignore the part of the polarimetric vector  
proportional to the $\rho$ (equivalently $\nu$) momentum in the $\tau$   
rest frame.   
We rely only on the part of the vector due to the differences of   
the $\pi^\pm$ and $\pi^0$ momenta, which manifests the spin state of the $\rho^\pm$.
  
In the Higgs rest frame the $\rho$ momentum represents a larger fraction of   
the Higgs's energy than the neutrino. Therefore, we abandon the   
reconstruction of the   
Higgs rest frame and instead we use the $\rho^{+}\rho^{-}$ rest   
frame which has the advantage that it is built only from directly visible decay products of the  
$\rho^{+}$ and $\rho^{-}$.\footnote{%
The use of correlation angles to measure the Higgs parity have already been
proposed, see, {\it e.g.} \cite{Nelson:1995vt}
or \cite{Grzadkowski:1995rx}, but their definitions depended on the $\tau$ and
Higgs (or $Z/\gamma$) rest frames which are difficult to measure. In our approach,
these frames are replaced by others which are easily measured and yet retain
significant sensitivity to the Higgs parity. In some special cases our approach
may open the way to studying the Higgs parity at hadron machines such as the
Tevatron or the LHC.
}   
  
We take for both $\rho$'s the  $\pi^{\pm} \pi^{0}$ decay channel.   
In the rest frame of the $\rho^{+}\rho^{-}$ system we define the acoplanarity angle,   
$\varphi^{*}$, between the two planes spanned by   
the immediate decay products (the $\pi^\pm$ and $\pi^0$) of the  
two $\rho$'s.   
  
\subsection{Defining an optimal variable}  
  
The variable $\varphi^{*}$ alone does not distinguish the scalar and psuedoscalar Higgs. To do  
this we must go further. The $\tau \to \pi^\pm \pi^0 \nu$ spin sensitivity   
is proportional to   
the energy difference  of  the charged and neutral pion  
(in the  $\tau$ rest frame), see  formula (\ref{formula}).   
We have to separate events into    
two zones, ${\cal C}$ and ${\cal D }$,  
  
\[  
{\cal C}:~~~~~~~~~~~~~~~~  y_1 y_2>0   
\]  
\[  
{\cal D}:~~~~~~~~~~~~~~~~  y_1 y_2<0    
\] where,  
\begin{equation}  
y_1={E_{\pi^{+}}-E_{\pi^{0}}\over E_{\pi^{+}}+E_{\pi^{0}}}~;~~~~~ y_2={E_{\pi^{-}}-E_{\pi^{0}}\over E_{\pi^{-}}+E_{\pi^{0}}}.  
\label{E-zone}   
\end{equation}\\  
$E_{\pi^{\pm}}$ and  $E_{\pi^{0}}$ are   
the $\pi^{\pm}, \pi^{0} $ energies in the respective   
$\tau^\pm$ rest frames.   
If the cuts are applied using the four-momenta from the generation   
level boosted to the $\tau$ rest frame without any smearing then we call them respectively   
${\cal C}_{bare}$ and ${\cal D}_{bare}$.  
If the cuts are applied using the smeared four-momenta boosted to the replacement $\tau^\pm$  
 rest frames (defined below), then they are called ${\cal C}_{reco}$ and  
${\cal D}_{reco}$.  
  
\begin{figure}[!ht]  
\setlength{\unitlength}{0.1mm}  
\begin{picture}(1600,800)  
\put( 375,750){\makebox(0,0)[b]{\large }}  
\put(1225,750){\makebox(0,0)[b]{\large }}  
\put(-260, -400){\makebox(0,0)[lb]{\epsfig{file=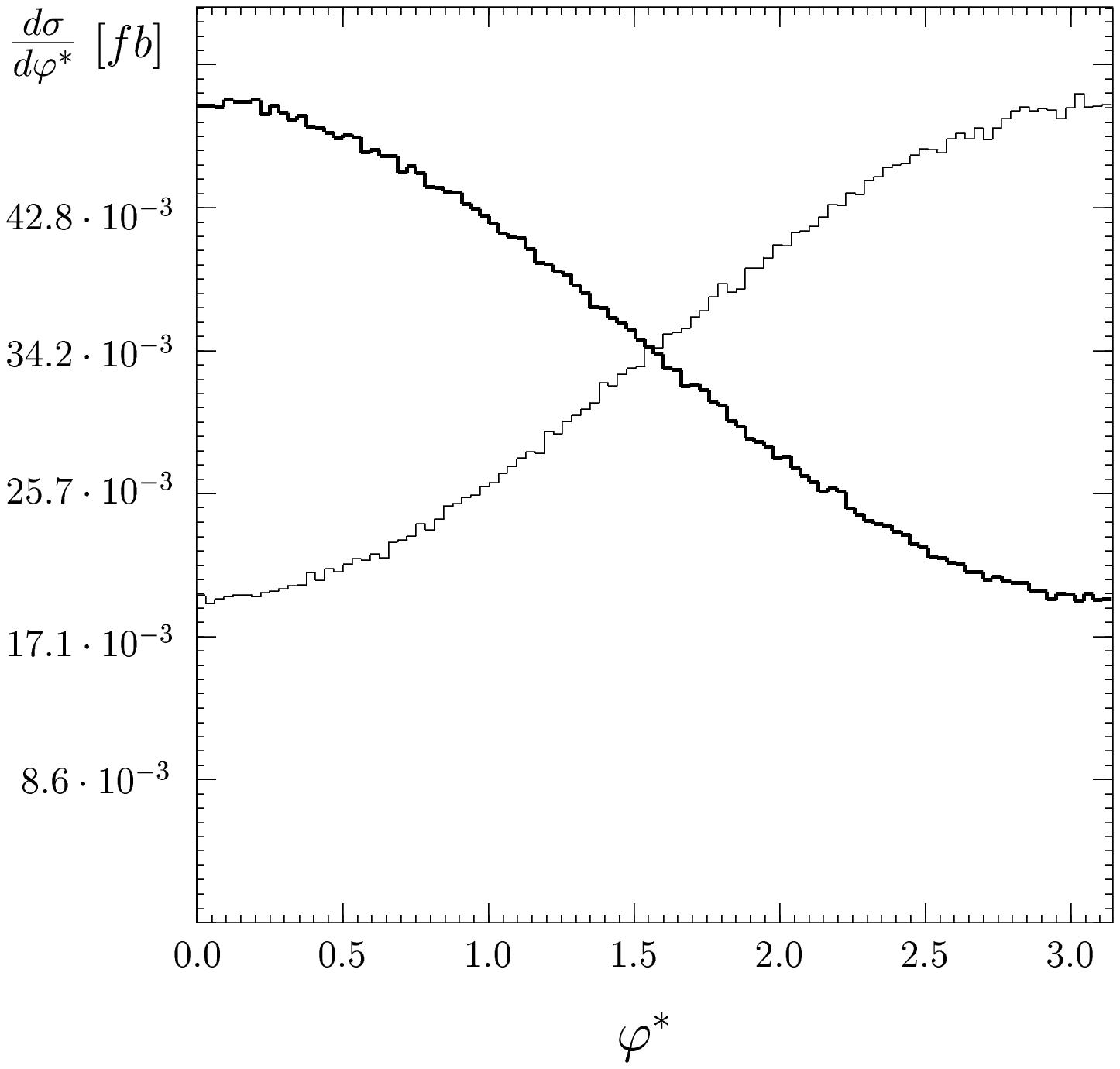,width=120mm,height=140mm}}}  
\put(580, -400){\makebox(0,0)[lb]{\epsfig{file=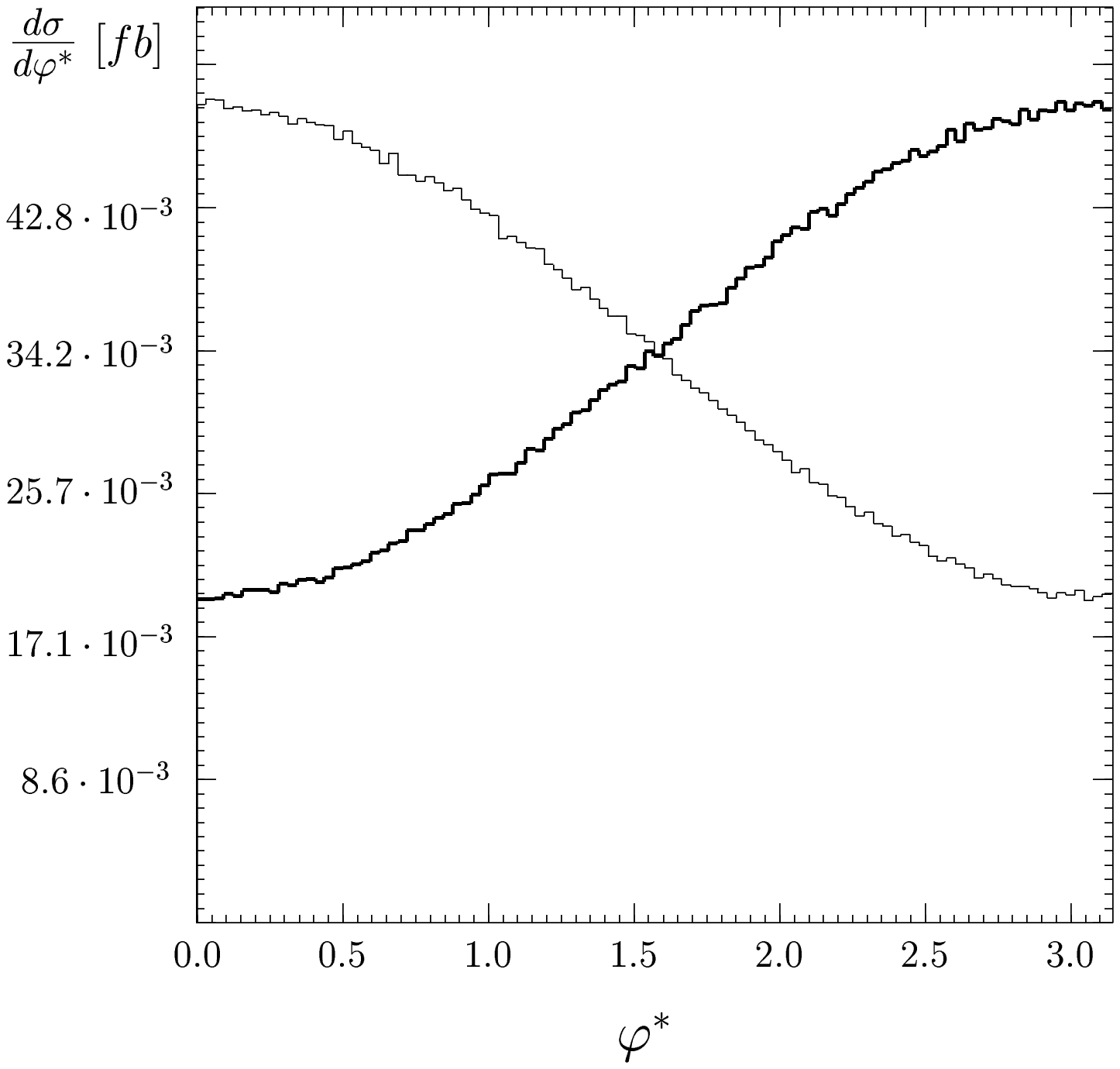,width=120mm,height=140mm}}}  
\end{picture}  
\caption  
{\it The $\rho^+ \rho^-$ decay products' acoplanarity distribution   
angle, $\varphi^*$,  
in the rest frame of the $\rho^+ \rho^-$ pair. A cut on the  
differences of the $\pi^\pm$ $\pi^0$ energies defined in their respective  
$\tau^\pm$ rest frames to be of  the same sign, selection ${\cal C}_{bare}$,   
is used in the left plot and the opposite   
sign, selection ${\cal D}_{bare}$, is used for the right plot. No smearing is done.  
 Thick lines denote the case   
of the scalar Higgs boson and thin lines the pseudoscalar one.   
Units valid for the  500 GeV $e^+e^-$ CMS  (scalar 120 GeV mass)
Higgsstrahlung production only. Otherwise arbitrary units.
}  
\label{aco}  
\end{figure}  

In Fig.~\ref{aco} we plot the distribution of $\varphi^{*}$,  
where the left hand plot contains the events where the  
energy difference between the $\pi^+$ and $\pi^0$ defined in the   
$\tau^+$  rest frame  
is of the same sign as the energy difference of $\pi^-$ and $\pi^0$  
 defined in $\tau^-$  rest frame (selection  ${\cal C}_{bare}$   
 and formula \ref{E-zone}).   
The right hand plot contains the events with the opposite signs for the two  
 energy differences (selection    
${\cal D}_{bare}$). It can be seen that the   
differences between   
the scalar and pseudoscalar Higgs boson are large.  
If the energy difference cut was not applied, we would have completely   
lost sensitivity to the Higgs boson parity.  
  
Unfortunately, since the $\tau$-lepton is not measurable,   
such a selection cut cannot be used directly.  
We now go on to define our choice for detection parameters.  
Then we will define  realizable methods for making the  energy cuts,  
${\cal C}_{reco}$ and ${\cal D}_{reco}$, and we will discuss phenomenologically   
sound results.  
  

\section{Detector Effects}     
   
To test the feasibility of the measurement, some assumptions about the detector   
effects have to be made.  
We include, as the most critical for our discussion, effects due to   
inaccuracies in the measurements of the  
$\pi$ momenta.  
We assume Gaussian spreads of the `measured' quantities with   
respect to the generated ones,  
and we use the following algorithm to reconstruct the energies of $\pi$'s    
in a measurable substitute for their respective $\tau^\pm$ rest frames.  
  
\begin{enumerate}  
\item   
{\bf Charged pion momentum:}  
 We assume a 0.1\% spread on its energy and direction.  
\item   
{\bf Neutral pion momentum:}  
We assume an energy spread of $ 5 \% \over \sqrt{E [GeV]}$. For the $\theta$ and $\phi$ angular   
spread we assume  $ {1 \over 3}  {2 \pi \over 1800}$. 
(These neutral pion resolutions can be   
achieved with a 15\% energy error and a 2$\pi$/1800 direction error in the gammas  
resulting from the $\pi^0$ decays.) \footnote{ We have studied a W/Si sampling EMCal 
with 1.25 m inner radius and
projective towers with 1800 segments in both $\theta$ and $\phi$ in a 5 Tesla field. The direction
of reconstructed photons can be defined by the center of energy of the hit cells in a cluster defined
by contiguous hit cells. The combination of the
small Moliere radius of tungsten, the fine segmentation and using detailed hit information yields a 
direction resolution of about 1/6 2$\pi$/1800 which is well below the resolution assumed above, see e.g. \cite{SantaC} and references therein for details.}

\item    
{\bf The replacement $\tau$ lepton rest frames and the $\pi^\pm$ and $\pi^0$ energy differences:}  
To make the measurement we replace the difficult to measure Higgs rest frame with the  
$\rho^+\rho^-$ rest frame. Define replacement four-monenta, $p_\pm$, 
in the $\rho^+\rho^-$ rest frame for the unmeasurable $\tau^\pm$ four-momenta:
 
a) Define $p^0_+ = p^0_- = m_H/2$.
 
b) Define the direction of the $p_+$ and $p_-$ three-momenta to be the direction of 
the $\rho_+$ and $\rho_-$ three-momenta in the $\rho^+\rho^-$ rest frame, respectively.

c) Define the magnitude of the $p_\pm$ three-momenta so that $p^2_+ = p^2_- = m^2_\tau$.

Boost the $\pi^+,\pi^0,\pi^-,\pi^0$ momenta to the respective rest frames, $p_+$ and $p_-$, of   
their replacement $\tau^+$ and $\tau^-$.
The $\pi$ energies defined this way are used in 
the ${\cal C}_{reco}$ and ${\cal D}_{reco}$ energy difference cuts.

\end{enumerate}

When  we  compare predictions for scalar and pseudoscalar, we should consider
not only properties of its decay, but
we should take care of the possible differences in the production mechanisms as well.
To avoid multitude of options    
we have excluded this point from our study. We  use  
Higgsstrahlung production mechanism  both for the scalar and   
pseudoscalar Higgs boson.  
Except for the size of the cross section, our analysis   
and its conclusions do not depend on the choice.  
We have checked that it is indeed the case by varying beam energies and choosing other   
production mechanisms \footnote{  
        Note that
	large Higgsstrahlung cross-section arise only if  Higgs boson has a sizable
        scalar component. In such a case  our method could
	measure its pseudoscalar admixture.
}.  
  
\subsection{The results of detector smearing}

\begin{figure}[!ht]  
\setlength{\unitlength}{0.1mm}  
\begin{picture}(1600,800)  
\put( 375,750){\makebox(0,0)[b]{\large }}  
\put(1225,750){\makebox(0,0)[b]{\large }}  
\put(-260, -400){\makebox(0,0)[lb]{\epsfig{file=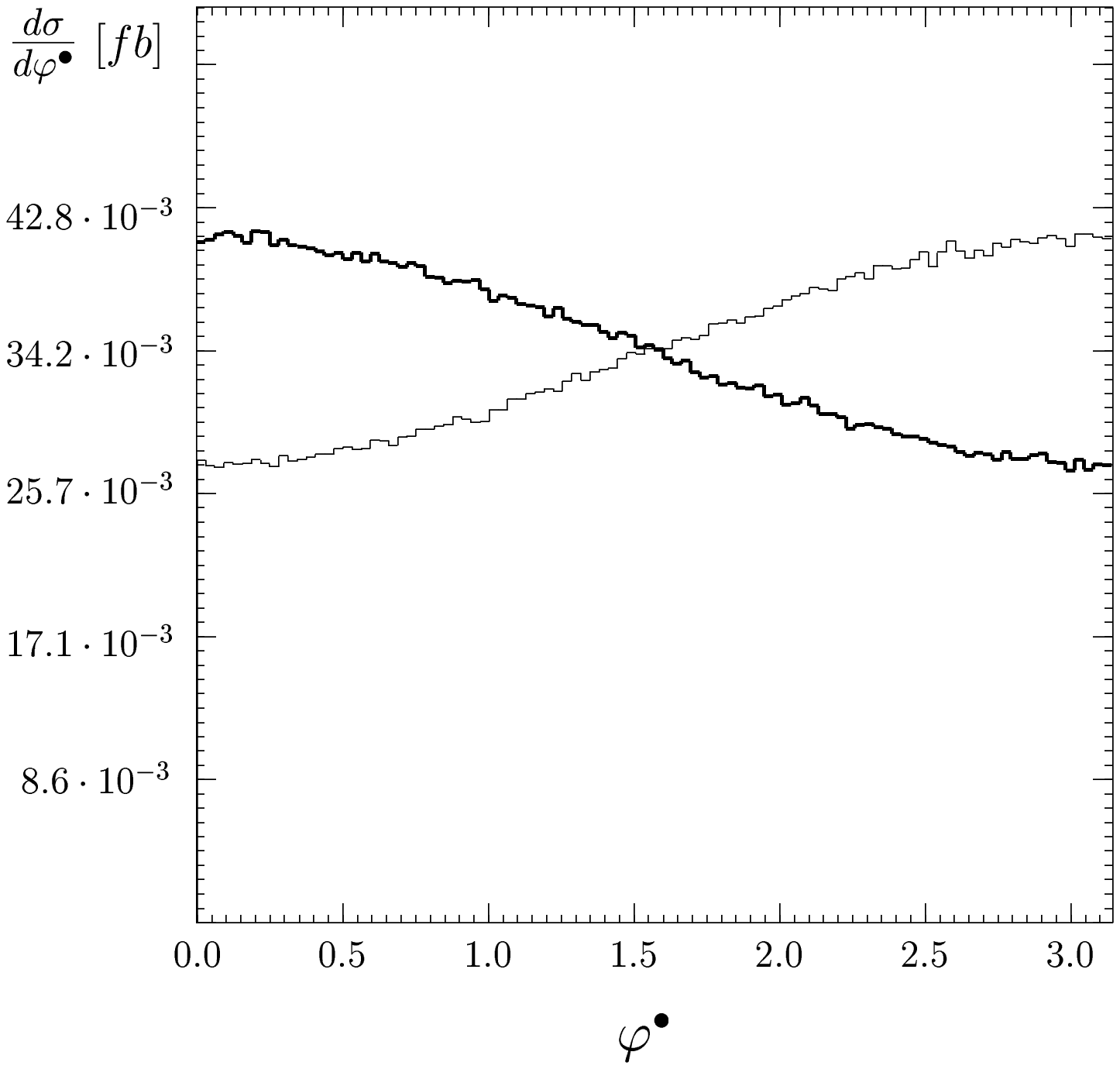,width=120mm,height=140mm}}}  
\put(580, -400){\makebox(0,0)[lb]{\epsfig{file=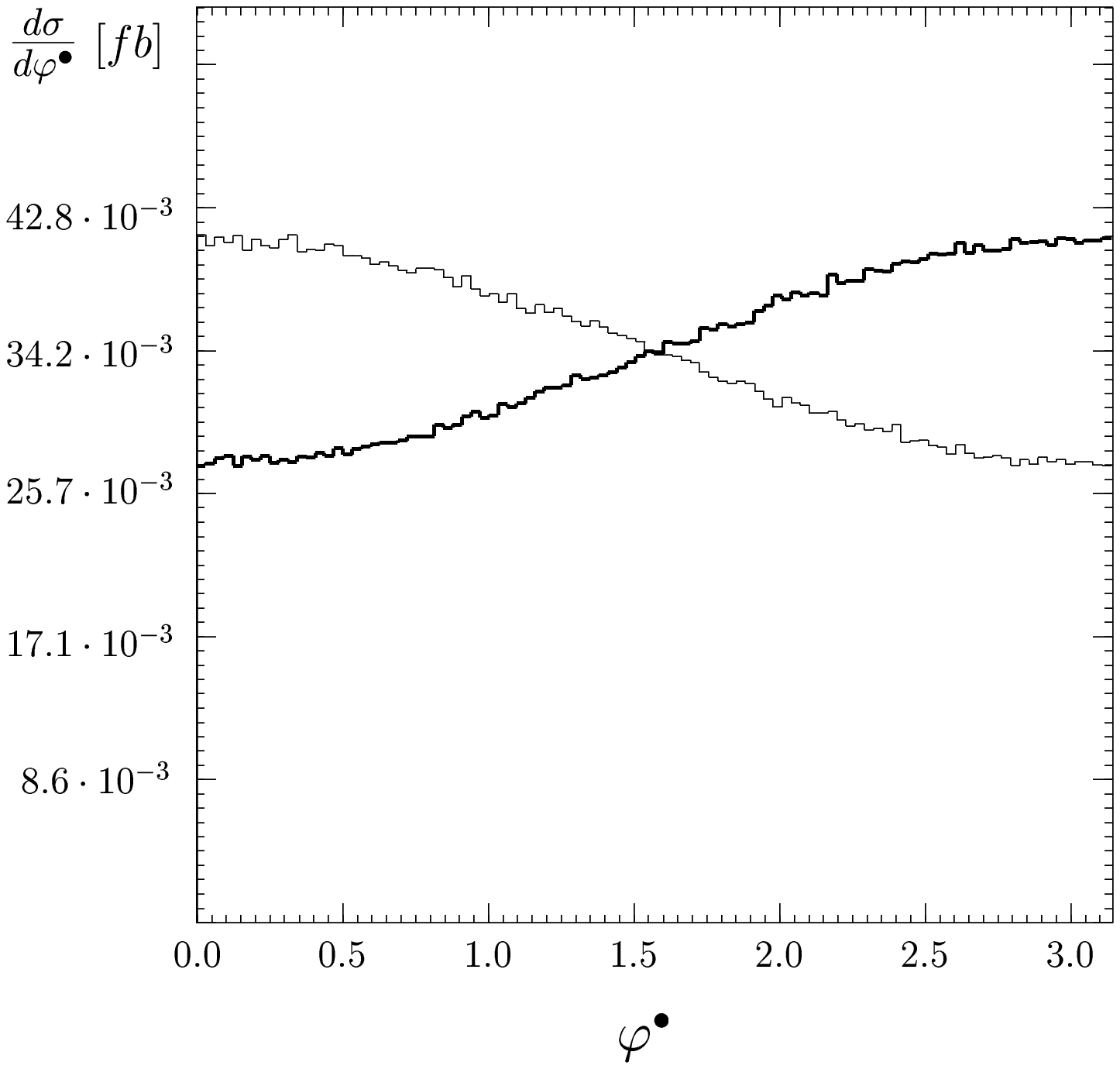,width=120mm,height=140mm}}}  
\end{picture}  
\caption  
{\it The $\rho^+ \rho^-$ decay products' acoplanarity distribution   
angle, $\varphi^\bullet$,  
in the rest frame of the $\rho^+ \rho^-$ pair. A cut on the  
differences of the $\pi^\pm$ $\pi^0$ energies defined in their respective replacement  
$\tau^\pm$ rest frames to be of  the same sign, selection ${\cal C}_{reco}$,   
is used in the left plot and the opposite   
sign, selection ${\cal D}_{reco}$, is used for the right plot. All smearing is included.  
 Thick lines denote the case   
of the scalar Higgs boson and thin lines the pseudoscalar one.  
Units valid for the  500 GeV $e^+e^-$ CMS  (scalar 120 GeV mass)
Higgsstrahlung production only. Otherwise arbitrary units.
}  
\label{acosm2}  
\end{figure}

If we use the true, generator level, $\tau$ rest frame to define   
the energy differences between the $\rho$ decay products   
but smear the momenta of the pions when used in the calculation of   
the acoplanarity angle, $\varphi^{\bullet}$, the resulting distributions   
(not shown) are very similar to the unsmeared   
case of Fig.~\ref{aco}.  
When we    
use the selection cuts ${\cal C}_{reco}$ and ${\cal D}_{reco}$  
(and thereby use the replacement $\tau$ rest frames as well as smearing   
the $\pi$ momenta) we obtain   
the results shown in Fig.~\ref{acosm2}. We see that the effects to be  
measured diminish but remain clearly visible \footnote{  
We have studied several options for the definition of the separation cuts   
 ${\cal C}_{reco}$ and ${\cal D}_{reco}$. In one case we have directly used the   
smeared laboratory frame energies.  In another, we have used all the information available from   
the reconstruction of the Higgs boson rest frame. All of these choices lead to practically   
identical versions of Fig.~\ref{acosm2}.  
}.  
  
\subsection{The potential measurement resolution}  
  
To determine the Higgs parity at an operating linear collider,   
a set of events will be collected over a period of time. An event selection  
will be made from the data set to isolate Higgsstrahlung events where the process  
$H/A \to \tau^{+}\tau^{-} \to \rho^{+}\bar{\nu}_{\tau}\rho^{-}\nu_{\tau}  
\to \pi^{+}\pi^0\bar{\nu}_{\tau}\pi^{-}\pi^0\nu_{\tau}$ occurred. This sample  
will be reconstructed and the distribution of the measured 
variable, $\varphi^{\bullet}$,   
defined in sections 3 and 5,  
will be compared with simulated reconstructed distributions for a scalar and a  
pseudoscalar Higgs such as those shown in Fig.~\ref{acosm2}. A goodness of fit  
test such as an unbinned maximum likelihood will be performed for both hypotheses.  
One hypothesis will be favored over the other and statistical techniques will   
be applied to estimate the confidence level of the conclusion.   
  
For present purposes, it  
is important to note that the level of certainty obtained will depend   
upon the specific sample collected. Another run under the same  
conditions with the same integrated luminosity would, in general, result in a  
sample of a different size and with different fit results due to the random   
sampling effects inherent in all measurements. Thus, we cannot a priori  
predict the level of certainty that will be achieved when the measurement   
is eventually made.  
However, we can simulate a large set of possible data samples that might be  
collected and gather some insight into the range of certainties possible.   
  
We have applied a binned maximum likelihood technique to a set of   
approximately 1300 simulated data samples where half the samples were derived  
from a ${\cal CP}$ even Higgs and half from a ${\cal CP}$ odd Higgs.   
The Higgs mass was taken to be 120 GeV, the beam energy was 250 GeV per beam.   
The $\rho^\pm$ were reconstructed from their gammas and $\pi^\pm$ decay products.   
The $\pi^\pm$ were smeared 0.1$\%$ in energy and direction, the gamma energy was smeared   
15\% and their direction by 2$\pi$/1800, a typical calorimeter cell size.   
We assumed an integrated luminosity of 500 $fb^{-1}$. Beamsstrahlung and ISR  
effects are included. To account for detector acceptance  
effects, event selection efficiency and impurity we assumed an overall efficiency of   
60\%. Studies \cite{evtsel1, evtsel2} have shown these are realistic estimates   
of event selection efficiencies and purities.  
  
Based on these assumptions, using a binned likelihood fit and comparing each data sample   
with the distributions for a ${\cal CP}$ even and a ${\cal CP}$ odd Higgs and comparing the resulting  
fits, we find that every data sample will identify the correct parent   
distribution with a confidence level of at least 95$\%$ and 86$\%$ of all samples will   
make the correct identification above the 3$\sigma$ confidence level. Thus, we see  
that we have an excellent chance of correctly determining the ${\cal CP}$ of the Higgs with  
this technique.

The technique is quite robust relative to the measurement resolutions of the charged pions and
the gammas (from the neutral pions). For example, decreasing the direction resolution of the gammas
even by a factor up to six has a negligible effect on the overall measurement resolutions cited above.
The $\varphi^{*}$ plots show that the odd and even states have on average
large angular differences, thus small errors in particle resolutions will not change $\varphi^{*}$
of an event enough, to make a significant impact on the overall distribution.

\section{Summary}  
  
We have studied the possibility of distinguishing   
a scalar from a pseudoscalar couplings of light Higgs   
to fermions  
using its decay to a pair of $\tau$ leptons  
and their subsequent decays to  
$\tau^\pm \to \nu_\tau \rho^\pm$ and $\rho^\pm \to \pi^\pm \pi^0$.  
  
We have discussed an observable which is very promising. Using reasonable
assumptions about the ${\cal SM }$ production cross section and about the measurement
resolutions
 we find that   
with 500 $fb^{-1}$ of luminosity at a 500 GeV $e^+e^-$ linear collider  
the ${\cal CP}$ of a 120 GeV Higgs can be measured to a confidence level greater   
than 95\%.  
To confirm the method we have used two distinct Monte   
Carlo programs \cite{Pandora} and \cite{Was2002,Golonka:2002iu} and the  
observables were coded independently. We emphasize that the technique  
is both model independent and independent of the Higgs production   
mechanism and depends only on good measurements of the Higgs decay products.  
Thus, this method may be applicable to other production modes including  
those available at proton colliders as well as at electron colliders.

Finally we note that several improvements are still possible.   
If, instead of unweighted events in our distributions as given in Fig. \ref{acosm2} we  use   
events weighted with $weight = |y_1 y_2| $ (see formula  \ref{E-zone})   
the statistical significance will increase%
\footnote{Instead of the weighted events we can fit 3 dimensional distributions  
spanned by the variables $\varphi^{*}$, $y_1$, $y_2$.}  
 by a factor of about 1.5. Other multi-meson final states such as $\tau \to \pi^+\pi^+\pi^- \nu$  
can be used to increase the statistical samples.   
As argued in \cite{Kuhn:1995nn} they should lead   
to additional data of  spin significance comparable to our  
 $\tau \to \rho \nu $ channel  
(if the appropriate observables are found).   
Finally, we may expect that a measurement of the   
$\tau$ flight direction may turn out to be helpful,   
similarly as it was the case for the $\tau$ polarization  
measurement at LEP 1 \cite{Heister:2001uh,Nikolic:1996gn}.    
For each new production mode that can be analyzed using this technique the  
sample size increases accordingly.  
We may expect the final   
statistical significance for the parity measurement to be at least factor of 2-3 times   
better than in our conclusions. Such improvements    
will  lead to more complex observables with many special cases   
and require good control of the systematic errors.   
We leave them to future studies but we think that, because of such   
possibilities, our estimation of the parity resolution will turn out   
to be largely conservative.

\subsection*{Acknowledgements}  
  
Discussions over the years with the experimental/phenomenological    
community analyzing   
$\tau$ lepton data  were essential for creating the necessary background to   
start this work and we acknowledge the work of people involved in the {\tt TAUOLA}   
project, in particular P. Golonka, S. Jadach, J. K\"uhn and E. Richter-W\c{a}s.  
M. Peskin has played an essential role in this project and we also thank D. Muller  
for helpful discussions.  
  

\begingroup\begin{thebibliography}{10}

\bibitem{Abe:2001gc}
{ACFA Linear Collider Working Group} Collaboration, {\it "Particle physics
  experiments at JLC"},
\href{http://arXiv.org/abs/hep-ph/0109166}{{\tt hep-ph/0109166}}.

\bibitem{:2001ve}
{NLC} Collaboration, {\it "2001 Report on the Next LinearCollider: A Report
  submitted to Snowmass 2001"}, \href{http://www.arXiv.org/abs/SLAC-R-571}{{\tt
  SLAC-R-571}}.

\bibitem{TESLA}
F.~Richard, J.~Schneider, D.~Trines, and A.~Wagner, {\it "TESLA Technical
  Design Report Part I: Executive Summary"},
  \href{http://www.arXiv.org/abs/hep-ph/0106314}{{\tt hep-ph/0106314}}.

\bibitem{Kramer:1994jn}
M.~Kramer, J.~H. K\"u{hn}, M.~L. Stong, and P.~M. Zerwas, {\em Z. Phys.} {\bf
  C64} (1994) 21,
\href{http://arXiv.org/abs/hep-ph/9404280}{{\tt hep-ph/9404280}}.

\bibitem{Barger1994}
V.~Barger, K.~Cheung, A.~Djouadi, B.~A. Kniehl, and P.~M. Zerwas, {\em Phys.
  Rev.} {\bf D49} (1994) 79--90.

\bibitem{Hagiwara1994}
K.~Hagiwara and M.~Stong, {\em Z. Phys.} {\bf C62} (1994) 99--108.

\bibitem{Hagiwara2000}
K.~Hagiwara, S.~Ishihara, J.~Kamoshita, and B.~Kniehl, {\em Eur. Phys. J.} {\bf
  C14} (2000) 457--468.

\bibitem{Grzadkowski:1995rx}
B.~Grzadkowski and J.~F. Gunion, {\em Phys. Lett.} {\bf B350} (1995) 218--224,
\href{http://arXiv.org/abs/hep-ph/9501339}{{\tt hep-ph/9501339}}.

\bibitem{Boe}
C.~A. Boe, M.~Ogreid, P.~Osland, and J.~Zhang, {\em Eur. Phys. J.} {\bf C9}
  (1999) 413,
\href{http://arXiv.org/abs/hep-ph/9811505}{{\tt hep-ph/9811505}}.

\bibitem{Skjold}
A.~Skjold and P.~Osland, {\em Phys. Lett.} {\bf B329} (1994) 305,
\href{http://arXiv.org/abs/hep-ph/9402358}{{\tt hep-ph/9402358}}.

\bibitem{Was2002}
Z.~W\c{a}s and M.~Worek, {\it "Transverse spin effects in $H/A\to \tau^+
  \tau^-;~ \tau^\pm \to \nu X^\pm$, Monte Carlo approach" },
  \href{http://www.arXiv.org/abs/hep-ph/0202007}{{\tt hep-ph/0202007}}.

\bibitem{Pierzchala:2001gc}
T.~Pierzcha\l{a}, E.~Richter-W\c{a}s, Z.~W\c{a}s, and M.~Worek, {\em Acta Phys.
  Polon.} {\bf B32} (2001) 1277--1296,
\href{http://arXiv.org/abs/hep-ph/0101311}{{\tt hep-ph/0101311}}.

\bibitem{Golonka:2002iu}
P.~Golonka, T.~Pierzcha\l{a}, E.~Richter-W\c{a}s, Z.~W\c{a}s, and M.~Worek,
enlarged version of the document {\tt hep-ph/0009302}, in preparation, to be
  submitted to {\it Comput. Phys. Commun.}

\bibitem{Bower}
G.~Bower, Talk given at the Chicago LC Workshop, 7-9 January 2002,
  transparencies available at:\\ {\tt
  http://needmore.physics.indiana.edu/\~{}rickv/nlc/chicago\_{}agenda.html}.

\bibitem{Pandora}
M.~Iwasaki and M.~E. Peskin, in preparation, see:\\ {\footnotesize \tt
  http://www-sldnt.slac.stanford.edu/nld/new/Docs/Generators/PANDORA$\_$PYTHIA%
.html}.

\bibitem{Pythia}
{T. Sjostrand} {\em et al.}, {\em Comput. Phys. Commun.} {\bf 135} (2001) 238.

\bibitem{jadach-was:1984}
S.~Jadach and Z.~W\c{a}s, {\em Acta Phys. Polon.} {\bf B15} (1984) 1151,
  \uppercase{E}rratum: {\bf B16} (1985) 483.

\bibitem{koralb:1985}
S.~Jadach and Z.~W\c{a}s, {\em Comput. Phys. Commun.} {\bf 36} (1985) 191.

\bibitem{Harton:1995dj}
J.~L. Harton, {\em Nucl. Phys. Proc. Suppl.} {\bf 40} (1995)
463--473.

\bibitem{Nelson:1995vt}
C.~A. Nelson, {\em Nucl. Phys. Proc. Suppl.} {\bf 40} (1995) 525--540,
\href{http://arXiv.org/abs/hep-ph/9411235}{{\tt hep-ph/9411235}}.

\bibitem{Jadach:1990mz}
S.~Jadach, J.~H. K\"{uhn}, and Z.~W\c{a}s, {\em Comput. Phys. Commun.} {\bf 64}
  (1990)
275.

\bibitem{Jezabek:1991qp}
M.~Je\.zabek, Z.~W\c{a}s, S.~Jadach, and J.~H. K\"{uhn}, {\em Comput. Phys.
  Commun.} {\bf 70} (1992)
69.

\bibitem{Jadach:1993hs}
S.~Jadach, Z.~W\c{a}s, R.~Decker, and J.~H. K\"{uhn}, {\em Comput. Phys.
  Commun.} {\bf 76} (1993)
361.

\bibitem{Barberio:1990ms}
E.~Barberio, B.~van Eijk, and Z.~W\c{a}s, {\em Comput. Phys. Commun.} {\bf 66}
  (1991)
115.

\bibitem{Barberio:1994qi}
E.~Barberio and Z.~W\c{a}s, {\em Comput. Phys. Commun.} {\bf 79} (1994)
291--308.

\bibitem{HERWIG}
G.~Corcella, I.~G. Knowles, G.~Marchesini, S.~Moretti, K.~Odagiri,
  P.~Richardson, M.~H. Seymour, and B.~R. Webber, {\em JHEP} {\bf 0101} (2001)
  010, \href{http://www.arXiv.org/abs/hep-ph/0011363}{{\tt hep-ph/0011363}}.

\bibitem{SantaC}
G.~R. Bower, {\it ``Cluser ID (a.k.a. Eflow)'' Talk presented
at the Santa Cruz LCD Workshop 28 June 2002.}  

\bibitem{evtsel1}
G.~R. Bower, {\it ``Proceedings of the Worldwide Study on Physics and
  Experiments with Future Linear $e^{+}e^{-}$ Colliders''}, (1999) 1049-1057.

\bibitem{evtsel2}
G.~R. Bower, {\it ``Physics and Experiments with Future Linear $e^{+}e^{-}$
  Colliders LCWS 2000''}, (2000) 920-923.

\bibitem{Kuhn:1995nn}
J.~H. K\"{uhn}, {\em Phys. Rev.} {\bf D52} (1995) 3128--3129,
\href{http://arXiv.org/abs/hep-ph/9505303}{{\tt hep-ph/9505303}}.

\bibitem{Heister:2001uh}
{ALEPH} Collaboration, A.~Heister {\em et al.}, {\em Eur. Phys. J.} {\bf C20}
  (2001) 401--430,
\href{http://arXiv.org/abs/hep-ex/0104038}{{\tt hep-ex/0104038}}.

\bibitem{Nikolic:1996gn}
I.~Nikolic, {\tt CERN-THESIS-99-031}.

\end{thebibliography}\endgroup
\providecommand{\href}[2]{#2}
\end{document}